\documentclass[useAMS,usenatbib,usegraphicx]{mn2e}
\usepackage{times}

\def\kms{$\rm km\, s^{-1}$}
\def\cm3{$\rm cm^{-3}$}

\def\n0{$\rm n_{0}$}
\def\B0{$\rm B_{0}$}

\def\mc{$\mu$m}

\def\L12{L$_{12\mu m}$~}
\def\F12{F$_{12\mu m}$~}

\def\Br{Br$\gamma$}
\def\fe2{[Fe\,{\sc ii}]}
\def\s3{[S{\sc iii}]}
\def\h2{H$_{2}$}
\def\w{$W_{\lambda}$}
\def\F{$F_{\lambda}$}

\def\pp{$\pm$}
\def\Sun{$_\odot$}

\title[Stellar Populations in the NIR]{The Stellar Populations of Starburst Galaxies Through
 near infrared spectroscopy}

\author[Riffel et al.]{R. Riffel$^{1}$\thanks{E-mail:
riffel@ufrgs.br}, M. G. Pastoriza$^{1}$, 
A. Rodr\'{\i}guez-Ardila$^2$\thanks{Visiting Astronomer at the Infrared Telescope Facility, which is operated by the University of Hawaii
under Cooperative Agreement no. NCC 5-538 with the National Aeronautics and Space Administration, Office of Space Science, Planetary Astronomy Program.}
and C. Maraston$^3$
\\$^{1}$Departamento de Astronomia, Universidade Federal do Rio Grande do Sul. Av. Bento Gon\c calves 9500, Porto Alegre, RS, Brazil.
\\$^2$ Laborat\'{o}rio Nacional de Astrof\'{i}sica/MCT - Rua dos Estados Unidos 154, Bairro das Nac\~oes.
\\$^3$ Institute of Cosmology and Gravitation, University of Portsmouth, Mercantile House, Hampshire Terrace, PO1 2EG, Portsmouth, United Kingdom}

\begin{document}

\date{}
\pagerange{\pageref{firstpage}--\pageref{lastpage}} \pubyear{2007}

\maketitle

\label{firstpage}

\begin{abstract}

We study the central (inner few hundred parsecs) stellar populations
of  four starburst galaxies (NGC34, NGC1614, NGC3310 and NGC7714) in the
near-infrared (NIR), from 0.8 to 2.4\mc, by fitting combinations of
stellar population models of various ages and metallicities. The
NIR spectra of these galaxies  feature many absorption
lines. For the first time, we fit simultaneously as much as 15 absorption features in the NIR. 
The observed spectra are best explained by stellar populations containing a
sizable amount ( 20 to 56 \% by mass) of $\sim$~1\,Gyr old stellar
population with Thermally Pulsing-Asymptotic Giant Branch stars. We found that the metallicity of
the stars which dominates the light is solar. Metallicities
substantially different from solar give a worse fit. Though the ages
and metallicities we estimate using the NIR spectroscopy are in agreement
with values from the literature based on the UV/optical, we find older
ages and a larger age spread. This may be due to the fact that the
optical is mostly sensitive to the last episode of star formation,
while the NIR better maintains the record of previous stellar
generations. Another interesting result is that the reddening
estimated from the whole NIR spectrum is considerably lower than that based on
emission lines. Finally, we find a good
agreement of the free emission line spectrum with photoionization
models, using as input spectral energy distribution the synthetic
composite template we derived as best-fit.
\end{abstract}

\section{Introduction}

Star-forming galaxies (starburst/H\,{\sc ii}) are among the best
laboratories to study the evolution of massive stars as well as the
physical processes that are associated with the earliest stages of
galaxy formation. The study of this kind of object starts with the
work of \citet{mgp67,mgp75} and \citet{sargent70}, where it is
suggested that some galaxies had their (nuclear) spectrum similar to
those of H\,{\sc ii} regions. Star-forming galaxies are easily
recognized from their proeminent emission line spectra, dominated by
hydrogen and helium recombination lines as well as strong forbidden
lines of sulfur, oxygen and nitrogen, among others. These sources
have also a prominent absorption line spectrum.

The Equivalent Widths (Ws) of the absorption features give insights on
the stellar populations of the host galaxy, their study is a critical
step in the analysis of the continuum and emission lines present in
the spectra of galaxies. Through the analysis of the stellar content
we get information about critical processes such as recent episodes of
star formation and the evolutionary history of the galaxy. 
The stellar populations of starburst galaxies and the central regions of Seyfert galaxies
have been mostly studied
in the UV and optical bands \citep[e.g.][and references therein]{bica88,schmitt96,rosa98,
raimann00a,raimann00b,charles2000,cid03,rosa04,westera04,cid05}.

The first use of the near infrared (NIR) dates back to nearly thirty
years ago. \citet{rieke80} employ NIR spectroscopy to 
study NGC\,253 and M\,82 and report the detection of strong  
CO band in these sources,  which suggests the presence of a large 
population of red giants and supergiants in the nuclear region of both sources. Since
the, other groups have used the near-IR to study star formation, 
in most cases the CO bands \citep[e.g.][and references therein]{orig93,oliva95,engelbracht98,lancon01} 
or photometric methods \citep[e.g.][]{moorwood82,hunt03}, but 
overall the whole NIR stellar spectrum (0.8 to 2.4\mc) with its many absorption features remains poorly explored.

The obvious reason to use the NIR for starburst galaxies is that this spectral range is
the most suitable one to unveil the stellar populations in highly obscured
sources. However, tracing star formation in the NIR is difficult \citep{origlia00}. 
To our knowledge, in this spectral region,
except for a few indicators such as the methods based on the CO(2-0)
first overtone or the Br$\gamma$ emission
\citep[e.g.][]{orig93,oliva95}, the detection of spectral features
allowing the identification and dating of stellar populations, mostly
those residing in the
inner few tens of parsecs of galaxies, remains an open question. As
stellar absorption features in the infrared are widely believed to
provide means for recognizing red supergiants \citep{oliva95}, they
arise as prime indicators for tracing starburst in galaxies.

Besides the  short-lived red supergiants typical of very
young stellar populations, the NIR hosts also the contribution
from the thermally- pulsating asymptotic giant branch (TP-AGB) stellar
phase, which is enhanced in young to intermediate age stellar
populations \citep[$0.2 \leq t \leq 2$ Gyr,][]{maraston98}. The
contribution of this stellar phase in stellar population models has
been recently included in both the energetics and the spectral
features by \citep[][hereafter M05]{maraston05}. In particular, these
models employ empirical spectra of oxygen-rich and carbon stars
\citep{lw00}, which are able to foresee characteristic NIR absorption
features such as the 1.1\mc\ CN band. TP-AGB stars leave a unique
fingerprint on the integrated spectra (cfr. M05, Fig. 14), hence when
detected they can help the age dating of stellar populations from
integrated light (M05). In fact, very recently \citet{riffel07}
detected the 1.1\mc\ CN band in the spectra of active galactic nuclei
and starburst galaxies. In addition, \citet{horacio05} propose that
carbon stars are natural candidates to explain the colour excess of
the nucleus with respect to the circumnuclear star formation ring of
NGC\,1241. Furthermore, the typical NIR spectral shape due to the
presence of TP-AGB stars has been detected in Spitzer data for
high-redshifted galaxies \citep{maraston06}.

The knowledge of the stellar populations that dominate the light in
the NIR becomes very important nowadays as the spectroscopic
observation of galaxies is being directed to this spectral region.
With the new generation of infrared arrays, it is now possible to
obtain spectra at moderate resolution on faint and extended
sources. Also, the availability of cross-dispersed spectrographs
offering simultaneous wavelength coverage in the interval
0.8$-$2.4~$\mu$m, allow the study of the NIR region avoiding the
aperture and seeing effects that usually affect {\it JHK} spectroscopy
done in long-slit mode and single-band observations.

With this in mind, we present for the first time  a detailed fitting 
of the whole 0.8-2.4\mc\ spectral range, for the
Starburst/H\,{\sc ii} galaxies studied by \citet[][hereafter
RRP06]{rrp06}.  This paper is structured as follows: The data
are presented in Sec.~\ref{data}. In Sec.~\ref{spmm} we describe the
fitting method and how the measurements are done. Results are presented and discussed in Sec.~\ref{results}. In
Sec.~\ref{gas} we focus on the free emission line spectrum. The final
remarks are given in Sec.~\ref{final}.

\section{The Data}\label{data}

\begin{figure*}
\centering
\includegraphics[width=14cm]{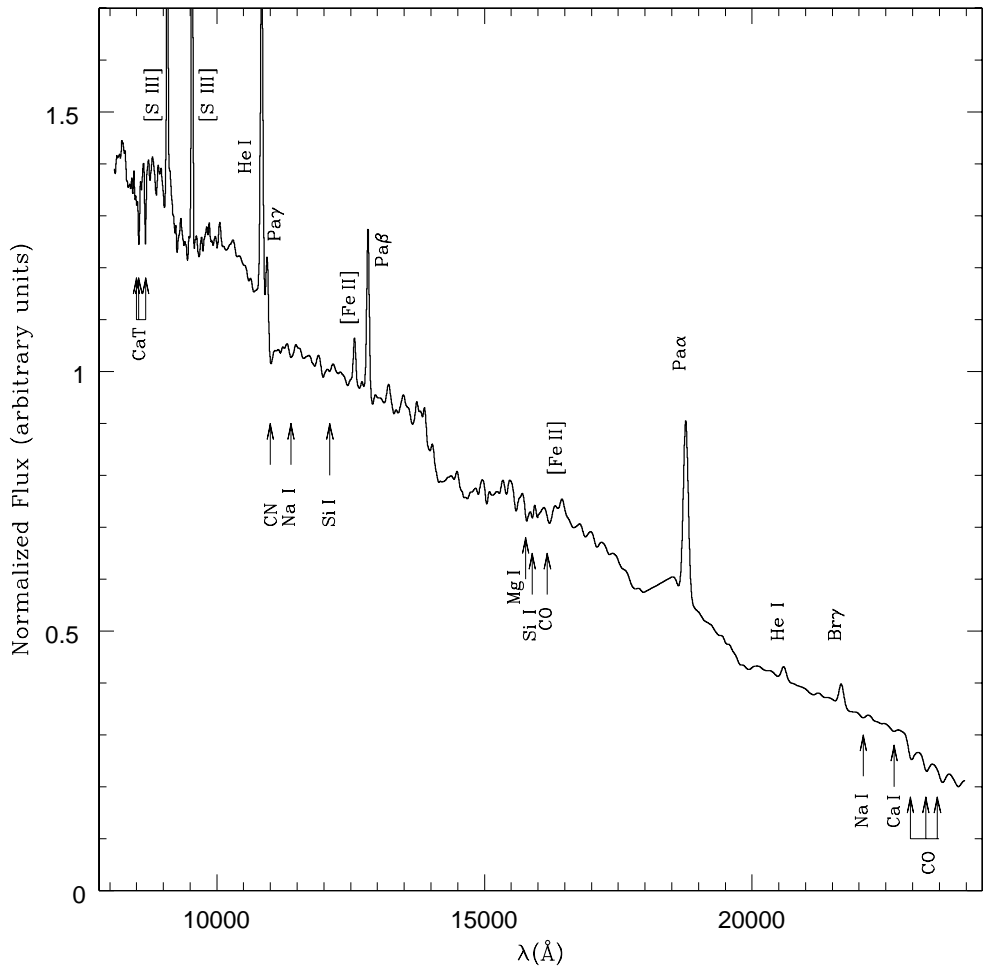}
\caption{The most prominent absorption and emission lines observed in the spectrum
of NGC\,7714. Note that here the spectral resolution was degraded in
order to match the M05 models (see Section \ref{spmm}).}
\label{abslines}
\end{figure*}

For this work we choose NGC\,34, NGC\,1614, NGC\,3310
and NGC\,7714, which are typical Starburst/H\,{\sc ii} galaxies and are
very well studied in other spectral ranges. The NIR spectra of these
sources,
presented in RRP06, were obtained at the NASA 3\,m Infrared Telescope
Facility (IRTF). The SpeX spectrograph \citep{ray03}, was used in the
short cross-dispersed mode (SXD, 0.8-2.4\mc). The detector used was a
1024$\times$1024 ALADDIN 3 InSb array with a spacial scale of
0.15''/pixel. A 0.8''$\times$15'' slit was employed giving a spectral
resolution of 360 \kms. For more details about instrumental
configuration and sample selection see RRP06. A rapid inspection of
Fig.\,13 in RRP06 allows us to the see that all the spectra are
dominated by strong absorption features due to stars. For reference,
we plot in Fig.~\ref{abslines} the most proeminent ones observed
in NGC~7714, including the
molecular bands of CN, CO and those of atomic transitions of Ca\,{\sc
ii}, Ca\,{\sc i}, Na\,{\sc i}, Si\,{\sc i} and Mg\,{\sc i}. Emission
lines present in the NIR spectrum of this object
are also identified.

\section{Analysis of the stellar populations: Method and Measures}\label{spmm}

In order to describe the age distributions of the stellar populations
of the sources and their metallicities, we compare the observed
spectra with various combinations of stellar population templates of
various ages and metallicities. In the analysis we use a method
similar to the one introduced by \citet{bica88}.  The latter employs
empirical star cluster spectra and combine them in order to estimate
the combination of ages and metallicities that allows the best match
to the galaxy spectra. Here instead, we use as input 
for the base of elements the synthetic
Simple Stellar Population (SSP) models computed by M05.

The fitting technique used here exploits the equivalent widths
($W_{\lambda}$) of spectral absorption features and the continuum
fluxes ($F_{\lambda}$) at selected wavelengths intervals, free from
emission/absorption lines. These quantities are measured in the galaxy
spectra and compared with combinations of the base of elements, which
in turn have known ages and metallicities. An algorithm generate these
combinations and compare the composite synthetic \w\ and $F_{\lambda}$
with the observed values. In practice, the algorithm tries to find the
minimum of the function,
\begin{equation}
f(x) {\rm =\sum_{i=1}^{N_W}[W_{gal}(i)-W_{SP}(i)]^2
+ \sum_{j=1}^{N_F}[F_{gal}(j)-F_{SP}(j)]^2}
\end{equation}
were $\rm W_{gal}$ and $\rm W_{SP}$ are, respectively, the equivalent
width of a given line {\it i} of the galaxy and of a Stellar
Population model (hereafter SP). The $\rm F_{gal}$ and $\rm F_{SP}$ are the continuum
fluxes at a selected point {\it j} of the galaxy and of a SP, respectively. The
possible solutions are those that reproduce, within allowed limits,
the \w\ and $F_{\lambda}$ observed in the galaxy spectra. The results
of this matching procedure are flux fractions at a
selected $\lambda$ of the different components of the base of elements.

In the present work we have upgraded the algorithm of
\citet{schmitt96} in order to include a search for the internal
reddening of the stellar population E(B-V)$\rm_{SP}$ for $F_{\lambda}$
points, free from emission/absorption lines, located in the NIR
region.  The \citet{ccm89} [CCM] reddening law was employed for this
purpose.

\begin{figure}
\centering
\includegraphics[width=8cm]{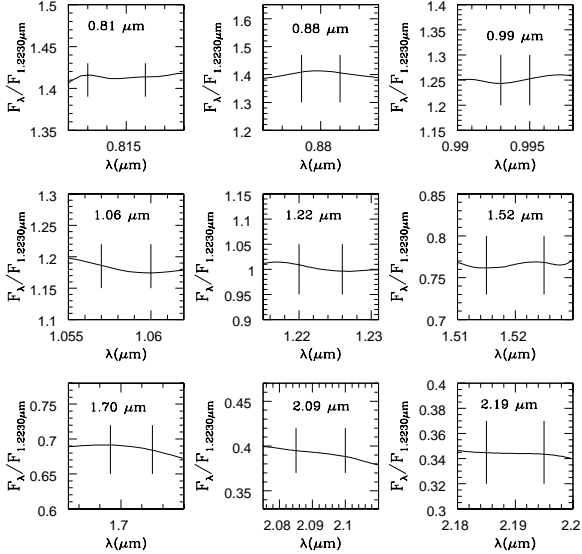}
\caption{Continuum intervals free from emission/absorption lines of
NGC\,7714. We use average values of the points between the
vertical lines.}
\label{contpp}
\end{figure}

We have defined the NIR continuum fluxes at 0.81\mc, 0.88\mc, 0.99\mc,
1.06\mc, 1.22\mc, 1.52\mc, 1.70\mc, 2.09\mc, and 2.19\mc, all 
normalized to 1.22\mc, to be used as input parameters for the
algorithm. As can be observed in Fig~\ref{contpp} we have used an
average value of the continuum points located between the vertical
lines. The measured values in the galaxy spectra as well as the
continuum intervals are presented in Tab~\ref{contflux}. The PACCE
\citep{tr07} code was used to measure the fluxes.

\begin{table}
\caption{Defined continuum fluxes, normalized to unity at 1.2230\mc, measured in our galaxy sample. 
The errors on \F\ are $\leq$ 2\% in all cases.}
\label{contflux}
\begin{tabular}{lcccc}
\noalign{\smallskip}
\hline
\hline
\noalign{\smallskip}
 Interval (\mc) &  NGC\,34 & NGC\,1614 & NGC\,3310 & NGC\,7714\\
\noalign{\smallskip}
\hline
\noalign{\smallskip}  
0.813-0.816 &	 0.86   &  0.76 &  1.29 & 1.41  \\
0.879-0.881 &	 0.94   &  0.88 &  1.27 & 1.41  \\
0.993-0.995 &	 1.02   &  1.00 &  1.19 & 1.25  \\
1.057-1.060 &	 1.03   &  1.03 &  1.13 & 1.18   \\
1.220-1.226 &	 1.00   &  1.00 &  1.00 & 1.00  \\
1.515-1.525 &	 0.91   &  0.85 &  0.79 & 0.77  \\
1.699-1.703 &	 0.92   &  0.84 &  0.72 & 0.69 \\
2.085-2.100 &	 0.63   &  0.54 &  0.41 & 0.39  \\
2.185-2.195 &	 0.56   &  0.49 &  0.37 & 0.34  \\
\noalign{\smallskip}
\hline
\noalign{\smallskip}
\end{tabular}
\end{table}

\begin{figure}
\centering
\includegraphics[width=8cm]{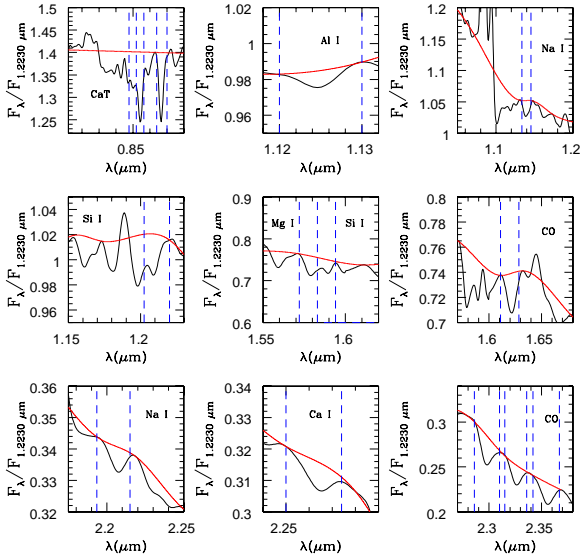}
\caption{Bandpasses (dashed lines) used to compute the equivalent widths of NGC 7714.
Al\,{\sc i} 1.1250$\mu$m and Si\,{\sc i} 1.2112$\mu$m are from NGC\,34. The red line (full line) 
is the ajusted continuum.}
\label{bandpass}
\end{figure}

Bandpasses and continuum points
used to compute the \w\ of the strongest absorption lines were taken from the literature or newly defined here. For the CaT lines and for the CN band we employed the
definitions of \citet{bica87} and \citet{riffel07}, respectively. The
\w\ were measured using the bandpass as listed in Tab.~\ref{eqw}.
For Al\,{\sc i}\,1.1285\,\mc, Na\,{\sc i}\,1.1383\,\mc\ and Si\,{\sc
i}\,1.2112\,\mc\ the continuum was adjusted as a cubic spline using
points free from emission/absorption lines between 1.02\,\mc\ and
1.23\,\mc. The continuum was also adjusted as a cubic spline for the
\w\ measurements of Mg\,{\sc i}, Mg\,{\sc i}\,1.5771\,\mc, Si\,{\sc
i}\,1.5894\,\mc\ and CO 1.6175\,\mc\ using points between 1.43\,\mc\
and 1.69\,\mc. In the same way, we have adjusted the continuum as a
cubic spline for the lines located in the $K$-band, using continuum
points free from contaminations of emission/absorption features in the
2.09-2.37\,\mc\ spectral range. In Fig.~\ref{bandpass} we illustrate
the continuum and bandpasses used to compute the $W_{\lambda}$. 

Note that the resolution of the observed spectra has been degraded
(see Fig.~\ref{abslines}) to that of the M05 models\footnote{20\,\AA\
at the bluest lambdas of our spectrum up to 1\,\mc, 50\,\AA\ from 1\,\mc\
to 1.6\,\mc\ and 100\,\AA\ up to the end of the galaxy
spectrum.}. This procedure was done using the task {\it gauss} of the
IRAF\footnote{IRAF is distributed by the National Optical Astronomy Observatories, which are operated by the Association of Universities for Research
in Astronomy, Inc., under cooperative agreement with the National
Science Foundation.} software. The measured values, as well as 
their uncertainties are listed 
in Tab.~\ref{eqw}. Errors were 
estimated using the standard deviation of three different choices of
the continuum level around a given $W_{\lambda}$, which basically
reflects the subjectiveness associated with continuum determinations.

\begin{table*}
\caption{Equivalent widths (in \AA) of the absorption bands measured on the sample galaxies.}
\label{eqw}
\begin{tabular}{lcccccc}
\noalign{\smallskip}
\hline
\hline
\noalign{\smallskip}
 Feature$\rm ^a$ & $\rm \lambda$(\mc) & Band Pass (\mc)& NGC\,34 & NGC\,1614 & NGC\,3310 & NGC\,7714\\
\noalign{\smallskip}
\hline
\noalign{\smallskip}  
CaT$\rm _1$ &  0.8498  &  0.8476-0.8520$\rm ^b$ & 3.75\pp0.06    &  4.57\pp0.07  &  2.44\pp0.21   & 2.36\pp0.02  \\
CaT$\rm _2$ &  0.8542  &  0.8520-0.8564$\rm ^b$ & 3.89\pp0.01    &  4.75\pp0.05  &  3.87\pp0.18   & 3.40\pp0.03  \\
CaT$\rm _3$ &  0.8670  &  0.8640-0.8700$\rm ^b$ & 3.70\pp0.16    &  4.05\pp0.09  &  3.12\pp0.16   & 3.31\pp0.13  \\
CN$\rm ^c$  &  1.0950  &  1.0780-1.1120         & 24.72\pp0.13   &  25.77\pp0.17 &  16.00\pp0.37  & 24.34\pp0.16 \\
Al\,{\sc i} &  1.1250  &  1.1200-1.1300         & 0.45\pp0.04    &      -        &     -	  &   - 	 \\
Na\,{\sc i} &  1.1395  &  1.1335-1.1455         & 1.55\pp0.10    &  2.76\pp0.08  &  3.10\pp0.11   & 1.16\pp0.17  \\
Si\,{\sc i} &  1.2112  &  1.2025-1.2200         & 3.57\pp0.50    &       -       &  	  -	  & 	 -	\\
Mg\,{\sc i} &  1.5771  &  1.5720-1.5830         & 5.64\pp0.30    &  6.34\pp0.40  &  5.05\pp0.09   & 4.42\pp0.04 \\
Si\,{\sc i} &  1.5894  &  1.5870-1.5940         & 1.93\pp0.03    &    -	         &  3.82\pp0.08   & 2.01\pp0.07  \\
CO	    &  1.6175  &  1.6110-1.6285         & 4.27\pp0.40    &  5.37\pp0.15  &  4.85\pp0.24   & 4.15\pp0.08  \\
Na\,{\sc i} &  2.2063  &  2.1936-2.2150         & 4.78\pp0.05    &  3.95\pp0.14  &  2.87\pp0.12   & 2.44\pp0.06  \\
Ca\,{\sc i} &  2.2655  &  2.2570-2.2740         & 4.48\pp0.04    &  4.07\pp0.02  &  2.36\pp0.13   & 3.57\pp0.01  \\
CO	    &  2.2980  &  2.2860-2.3100         & 16.72\pp0.29   &  17.66\pp0.45 &  13.87\pp0.44  & 14.22\pp0.32 \\
CO	    &  2.3255  &  2.3150-2.3360         & 8.91\pp0.14    &  9.75\pp0.16  &  9.06\pp0.35   & 9.24\pp0.13  \\
CO	    &  2.3545  &  2.3420-2.3670         & 14.68\pp0.50   &  16.02\pp0.15 &  15.74\pp0.12  & 13.20\pp0.04 \\
\noalign{\smallskip}
\hline
\noalign{\smallskip}
\multicolumn{6}{l}{a) The identifications of the atomic lines are based on}\\
\multicolumn{6}{l}{ \citet{cus05}.} \\
\multicolumn{6}{l}{b) From \citet{bica87}.} \\
\multicolumn{6}{l}{c) For details see  \citet{riffel07}.} \\
\end{tabular}
\end{table*}

The spectral base was taken from the evolutionary population synthesis (EPS) models of M05. These models are
particularly interesting for studying the stellar populations in the
NIR because, as discussed in the Introduction, they include empirical
spectra of carbon and oxygen rich stars evolving through the TP-AGB
phase. Thus, the models can predict the strength of molecular
features like CH, CN and C$\rm _2$. \citet{riffel07} report, for 
the first time, the detection of the
1.1\mc\ CN band in a sample of galaxies including the sources
studied in the present work. CN is particularly strong in young/intermediate
stellar populations with ages in the range $\sim$0.3$\leq t \leq$2\,Gyr,  though 
is also detected in red supergiants of massive stars \citep[e.g.][]{lw00,lancon07}.

\begin{figure*}
\centering
\includegraphics[width=14cm]{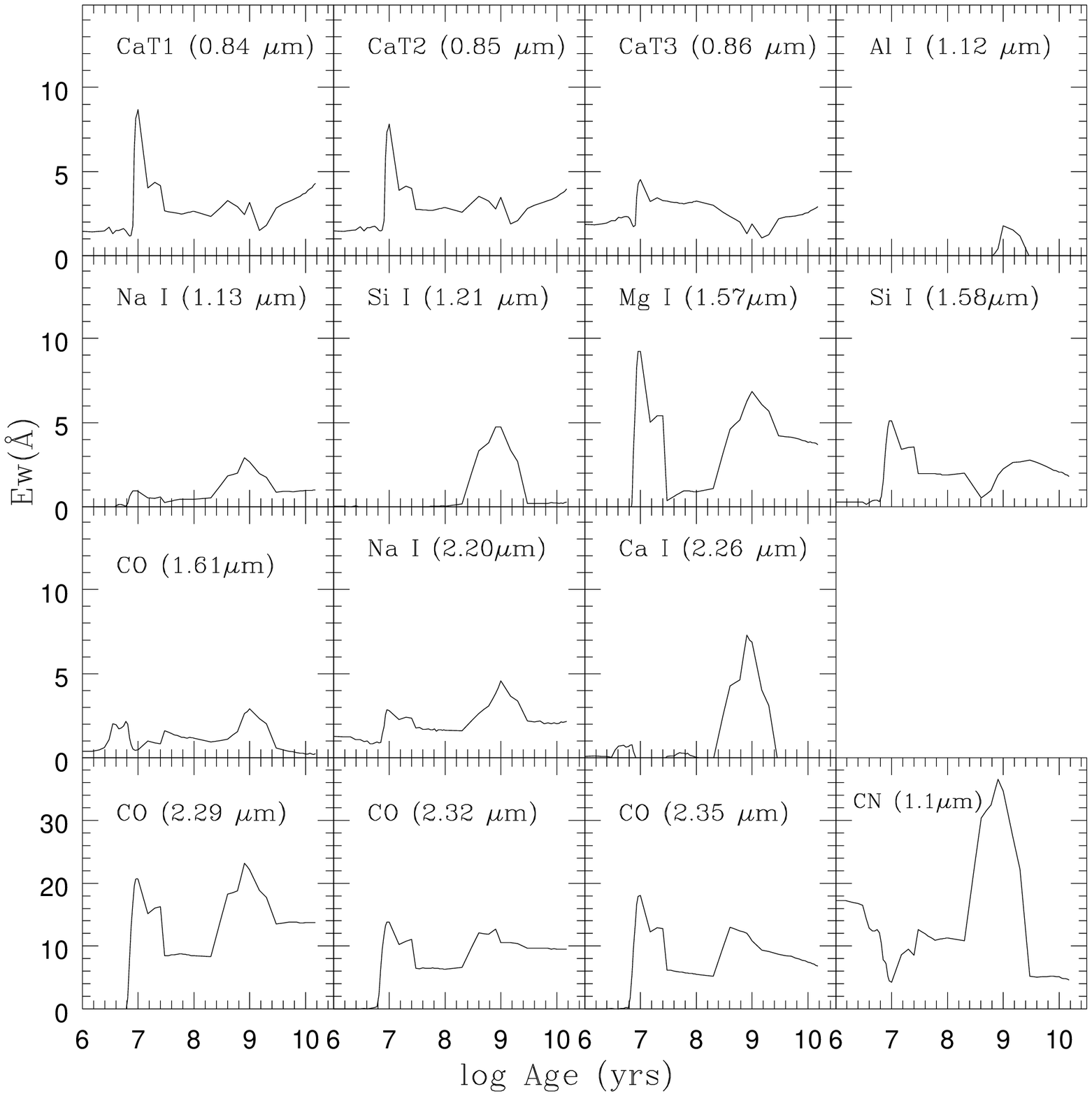}
\caption{Absorption lines as functions of age. Note that the 2.3\mc\ CO bands
as well as the CN are plotted with a different scale for \w.}
\label{linesage}
\end{figure*}

As base set we consider solar metallicity SSPs with a Salpeter initial
mass function and the following age distributions: 0.001, 0.01, 0.03, 0.05 0.2,
0.5, 0.7, 1 and 13\,Gyr. The ages are chosen taking into account the following:
({\it i}) very young starbursts (1\,Myr population) can be present; ({\it ii}) 
red supergiants (RSG) are important contributors to the NIR 
\citep[e.g.][and are included in the 10, 30 and 50 Myr populations 
following the input stellar tracks of the M05 models (from the Geneva 
database, see M05 for references) ]{oliva95,lancon01}; ({\it iii}) stars
in the TP-AGB phase \citep[M05,][ populations 
with 200, 500, 700 Myr and 1\,Gyr]{riffel07}. We additionally include an old model 13\,Gyr.
The behavior of the \w\ with age is presented in Fig.~\ref{linesage}.  
The base set is capable to discriminate between the stellar populations 
in the RSG and TP-AGB phases (see Fig.~\ref{linesage}). Here it should be noted that while 
the M05 models include empirical spectra of C-O-rich TP-AGB stars, they do not 
include empirical spectra of red supergiants. The latter show a large scatter 
in observed properties \citep{lancon07}. In addition, also the energetics of the RSG 
phase is not well known as it depends on mass-loss in massive stars, which is in turn 
highly uncertain. On the other hand, RSG display near-IR absorption bands and a 
proper spectrum should be used. To explore how much we could be in 
error by neglecting the true spectrum of a RSG, we made a 30 Myr, M05 
model in which we include a stellar spectrum of an O-rich star from 
the database of \citet{lw00}, that displays very similar 
features to a RSG \citep[see][Fig. 7]{lw00}. We have then 
repeated the fitting procedure for the galaxy NGC\,7714 using this 30 Myr 
model instead of the standard one. We find that the results are essentially the same.
The measured \w\ and $F_{\lambda}$ for the SSPs used in the base are given in Tab.~\ref{eqw_base} 
and Tab~\ref{flux_base}.

It is important to note that we have tested other metallicities in the range $\frac{1}{50}\,Z_{\odot}
\leq Z \leq 2\,Z_{\odot}$. To find the best solution we compute the $\chi$ of the 
residuals of the \w\ and \F\ values, i.e. we solve the equations: 

\begin{equation}
\chi(W_{\lambda})=\frac{\left[\displaystyle\sum_{i=1}^{i=N_W}\Delta W_{\lambda}(i)^2\right]^{\frac{1}{2}}}{N_W}
\end{equation}

and

\begin{equation}
\chi(F_{\lambda})=\frac{\left[\displaystyle\sum_{j=1}^{j=N_F}\Delta F_{\lambda}(j)^2\right]^{\frac{1}{2}}}{N_F}.
\end{equation}

The $\Delta W_{\lambda}$ and $\Delta F_{\lambda}$ are the differences between the observed and synthetic
values for the \w\ and \F, respectively. As can be seen in Tab.~\ref{chisqr} we find that the solar
metallicity provides the best solution for the four galaxies. This
result is in agreement with the analysis of the
optical region of spiral bulges \citep[e.g.][]{schmitt96,saraiva01}. 
Hence, as a first result of this paper, we find that the nuclear
regions of these starburst galaxies are characterized by a solar-like metallicity.

\begin{table}
\begin{tiny}
\renewcommand{\tabcolsep}{0.70mm}
\caption{Equivalent widths (in \AA) for the SSPs used in the base.}
\label{eqw_base}
\begin{tabular}{lccccccccccccccc}
\noalign{\smallskip}
\hline
\hline
\noalign{\smallskip}
Age$^a$ &CaT$\rm _1$ & CaT$\rm _2$ & CaT$\rm _3$& CN      & Al\,{\sc i} & Na\,{\sc i} & Si\,{\sc i} & Mg\,{\sc i} &  Si\,{\sc i}   &   CO   &Na\,{\sc i} & Ca\,{\sc i} & CO     & CO     &   CO \\
\noalign{\smallskip}
\hline
\noalign{\smallskip} 
   0.001 &    1.45  &	     1.49   &	   1.88   & 0.00  & 0.00	 &     0.00    &     0.00     &     0.00    &	   0.30      & 0.00  & 1.29	  &	0.00   &   0.00  & 0.00   & 0.00 \\ 
   0.01	 &    8.70  &	     7.82   &	   4.53   & 0.00  & 0.00	 &     1.15    &     0.00     &     9.24    &	   5.12      & 0.74  & 2.81	  &	0.00   &   20.66 & 13.81  & 18.02 \\ 
   0.03  &    2.70  &        2.81   &      3.35   & 12.52 & 0.00         &     0.41    &     0.00     &     0.25    &      1.81      &  1.06 & 1.60       &     0.00   &   7.53  & 6.13   & 6.33  \\ 
   0.05  &    2.56  &        2.75   &      3.20   & 11.26 & 0.00         &     0.57    &     0.00     &     0.62    &      1.81      &  0.84 & 1.59       &     0.00   &   7.79  & 6.20   & 6.07  \\ 
   0.2 	 &    2.33  &	     2.59   &	   3.00   & 10.84 & 0.00	 &     0.71    &     0.00     &     1.10    &	   2.01      & 0.68  & 1.62	  &	0.00   &   8.30  & 6.59   & 5.25 \\ 
   0.5	 &    3.08  &	     3.38   &	   2.16   & 31.50 & 0.00	 &     2.25    &     3.59     &     4.91    &	   0.77      & 2.98  & 2.88	  &	4.91   &   18.51 & 11.97  & 12.69 \\ 
   0.7 	 &    2.67  &	     3.00   &	   1.64   & 34.63 & 0.00	 &     2.89    &     4.31     &     5.77    &	   1.49      & 3.76  & 3.47	  &	6.70   &   21.10 & 12.33  & 12.22 \\ 
   1  	 &    3.16  &	     3.48   &	   1.91   & 34.65 & 1.64	 &     3.07    &     4.74     &     6.86    &	   2.26      & 4.40  & 4.58	  &	7.55   &   22.13 & 10.54  & 10.87 \\ 
  13 	 &    4.07  &	     3.76   &	   2.78   & 0.00  & 0.00	 &     1.24    &     0.22     &     3.80    &	   1.92      & 0.63  & 2.12	  &	0.00   &   13.71 & 9.51   & 7.10 \\ 
\noalign{\smallskip}
\hline
\noalign{\smallskip}
\multicolumn{6}{l}{a) In Gyr.} \\
\end{tabular}
\end{tiny}
\end{table}

\begin{table}
\begin{scriptsize}
\caption{Continuum fluxes, normalized to unity at 1.2230\mc, measured on the base SSPs.}
\label{flux_base}
\begin{tabular}{lccccccccc}
\noalign{\smallskip}
\hline
\hline
\noalign{\smallskip}
Age         &    0.81  & 0.88  & 0.99   & 1.06   & 1.22   & 1.52    &  1.70 & 2.09   &  2.19 \\
\noalign{\smallskip}
\hline
\noalign{\smallskip}  
 0.001      &	4.57 & 3.48 & 2.21 & 1.74 & 1.00 & 0.43 & 0.28 & 0.28 & 0.10 \\ 
 0.01	    &	1.40 & 1.50 & 1.49 & 1.32 & 1.00 & 0.81 & 0.76 & 0.76 & 0.35 \\ 
 0.03	    &   2.57 & 2.36 & 1.75 & 1.47 & 1.00 & 0.57 & 0.44 & 0.21 & 0.18  \\ 
 0.05	    &   2.42 & 2.18 & 1.67 & 1.43 & 1.00 & 0.60 & 0.46 & 0.23 & 0.19  \\ 
 0.2	    &	2.21 & 1.98 & 1.58 & 1.38 & 1.00 & 0.62 & 0.47 & 0.47 & 0.20 \\ 
 0.5	    &	1.50 & 1.51 & 1.28 & 1.24 & 1.00 & 0.69 & 0.68 & 0.68 & 0.33 \\ 
 0.7	    &	1.47 & 1.49 & 1.28 & 1.24 & 1.00 & 0.70 & 0.71 & 0.71 & 0.35 \\ 
 1	    &	1.45 & 1.49 & 1.29 & 1.24 & 1.00 & 0.71 & 0.72 & 0.72 & 0.36 \\ 
 13	    &	1.58 & 1.50 & 1.37 & 1.26 & 1.00 & 0.74 & 0.62 & 0.62 & 0.28 \\ 
\noalign{\smallskip}
\hline
\noalign{\smallskip}
\end{tabular}
\end{scriptsize}
\end{table}

\begin{table}
\renewcommand{\tabcolsep}{0.70mm}
\caption{$\chi$ values for the different metallicities.}
\label{chisqr}
\begin{tabular}{lcccccccccc}
\hline
\hline
\noalign{\smallskip}
                      &  \multicolumn{5}{c}{$\chi$(\w)} &  \multicolumn{5}{c}{$\chi$(\F)}  \\
\cline{3-6}
\cline{8-11}
\noalign{\smallskip}
Source&             & 2\,$Z$\Sun & $Z$\Sun & $\frac{1}{2}Z$\Sun &  $\frac{1}{50}Z$\Sun & & 2\,$Z$\Sun & $Z$\Sun & $\frac{1}{2}Z$\Sun &  $\frac{1}{50}Z$\Sun\\
\noalign{\smallskip}
\hline
NGC\,34             & & 2.26        &   1.08  &     1.39           &     1.69        & &    0.21  & 0.10  & 0.22 & 0.23  \\ 
NGC\,1614           & & 2.90        &   1.53  &     2.28           &     2.14        & &    0.11  & 0.03  & 0.11 & 0.18  \\  
NGC\,3310           & & 2.22        &   1.54  &     1.72           &     2.31        & &    0.06  & 0.04  & 0.08 & 0.07   \\ 
NGC\,7714           & & 2.13        &   1.35  &     1.40           &     5.82        & &    0.03  & 0.02  & 0.04 & 0.04   \\  
\noalign{\smallskip}  
\noalign{\smallskip}
\hline
\noalign{\smallskip}
\end{tabular}
\end{table}

\section{Results}\label{results}

The results of our fitting procedure, in terms of flux
fractions at 1.2230\,\mc, are shown in Fig~\ref{popspec} and
Tab.~\ref{sps}.  As can be seen, there is a good
agreement between the observed spectrum and the synthetic one
for all galaxies.  We also list the derived E(B-V) as well as the 
morphological type of the sources, as obtained from NED. The stellar population in terms of mass fraction was
calculated using the mass-to-light ratio (M/L) in the
$J$-band~\footnote{see www.dsg.port.ac.uk/~marastonc/}, assuming that
the $J$-band (centred at $\sim$1.2\mc) M/L is equal to
M/L$_{1.2230\mu\rm m }$, where L$_{1.2230\mu\rm m }$  is the continuum 
luminosity at 1.2230\mc. The mass fractions  are listed in
Tab.~\ref{sps}. We compare the flux and mass fractions of each
population for the four sources in Fig.~\ref{pop}.  
In the remaining of this section we comment on the results for each 
individual source.

\begin{table*}
\renewcommand{\tabcolsep}{0.70mm}
\caption{Results in terms of flux and mass fractions (first and second line, respectively).}
\label{sps}
\begin{tabular}{lccccccccccc}
\noalign{\smallskip}
\hline
\hline
\noalign{\smallskip}
 Source & Morphology$^{\dagger}$&E(B-V) & \multicolumn{9}{c}{ \% of age contributions$^{\ddagger}$.}  \\
 \noalign{\smallskip}
\cline{4-12}
\noalign{\smallskip}
&                     & (mag) & 0.001 & 0.01& 0.03  & 0.05& 0.2 & 0.5 & 0.7 & 1   & 13 \\
\noalign{\smallskip}
\hline
NGC\,34    & Sc       & 0.95\pp0.05  & 1\pp2 & 5\pp5  &  8\pp7   &  7\pp7   &  7\pp7   & 6\pp7 & 10\pp9  & 53\pp7  & 3\pp2     \\ 
	   &	      &              &1\pp4  & 0\pp1  &  0\pp1   &   0\pp1  &  7\pp14  & 4\pp10& 8\pp15  & 56\pp14 & 23\pp31  \\
 NGC\,1614 & SB(s)c   & 1.00\pp0.06  & 1\pp1 & 1\pp4  &  6\pp2   & 12\pp3   &  6\pp3   & 4\pp8 &  6\pp5  & 43\pp7  & 21\pp8   \\  
	   &	      &              &0\pp1  & 0\pp0  &  0\pp0   &   0\pp0  &  3\pp3   & 1\pp7 & 2\pp5   & 20\pp9  & 73\pp72  \\
 NGC\,3310 & SAB(r)bc & 0.55\pp0.03  & 2\pp1 &12\pp6  & 13\pp10  & 13\pp11  & 12\pp11  & 4\pp5 &  6\pp7  & 30\pp6  & 8\pp4    \\ 
	   &	      &              &1\pp1  & 1\pp0  &  1\pp0   &   1\pp0  &  10\pp16 & 2\pp5 & 4\pp9   & 27\pp9  & 54\pp45  \\
 NGC\,7714 & SB(s)b   & 0.46\pp0.02  & 7\pp3 & 7\pp4  & 10\pp8   & 13\pp10  & 11\pp9   & 4\pp4 &  6\pp5  & 34\pp4  & 8\pp3    \\  
 	   &	      &              &3\pp6  & 0\pp0  &  0\pp0   &   0\pp0  &  9\pp16  & 2\pp5 & 4\pp8   & 30\pp8  & 51\pp42  \\
\noalign{\smallskip}  
\noalign{\smallskip}
\hline
\noalign{\smallskip}
\multicolumn{9}{l}{$\dagger$ From NASA/IPAC Extragalactic Database.} \\
\multicolumn{6}{l}{$\ddagger$ Ages in Gyr.} \\
\end{tabular}
\end{table*}

\subsection{NGC\,34}

\begin{figure*}
\centering
\resizebox{\hsize}{!}{\includegraphics*{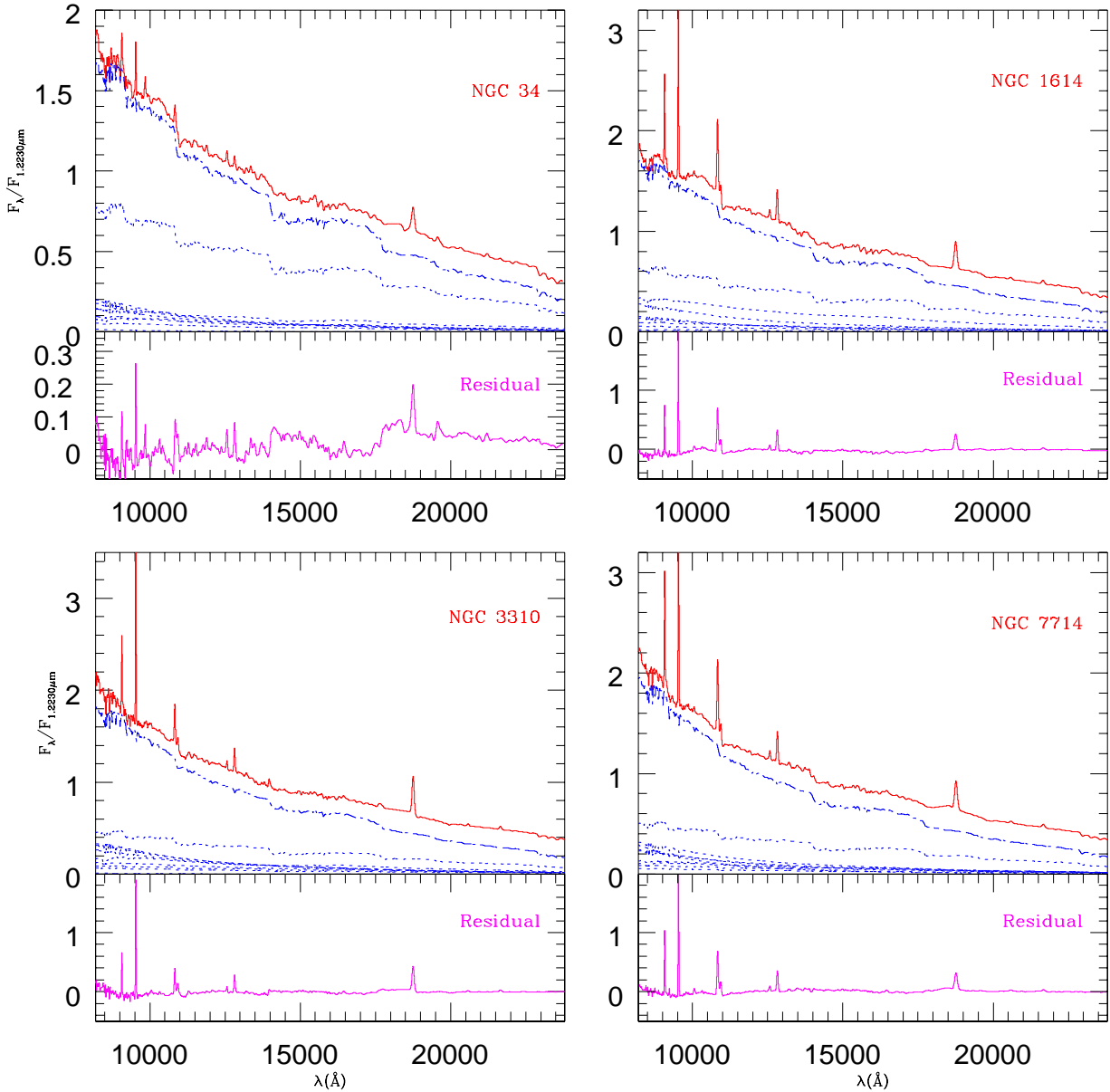}} 
\caption{Fitting results in terms of flux
contribution at 1.2230\mc. The galaxies are identified. {\it Top
panel:} solid lines represent the observed spectra, shifted by a
constant and smoothed to the models resolution. Dashed lines represent
the sum of the different stellar populations contributions. Dotted
lines show the component stellar populations, according
to Tab.~\ref{sps}. {\it Bottom panel:} the synthetic template is subtracted
from the galaxy observed spectrum.}
\label{popspec}
\end{figure*}

\begin{figure*}
\centering
\includegraphics[width=13cm]{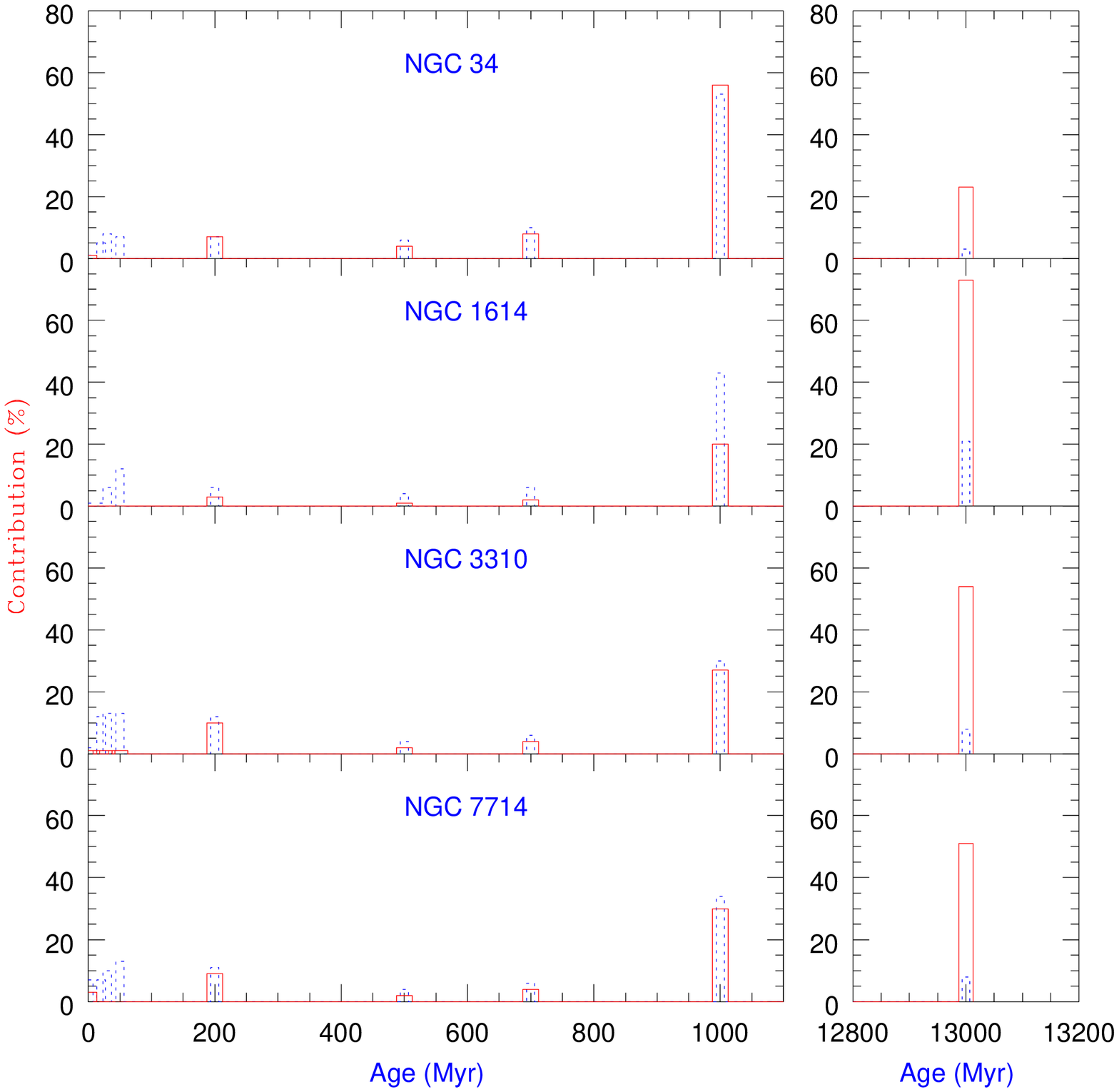}
\caption{Percentage contributions of each SSP to the total stellar
population of the galaxies. Solid lines represent the mass contribution
whereas dashed lines the flux fraction.  Note that the region between 12800 and 13200
Myr is shown in a separate box (right panel) for displaying purposes, as well as 
that this figure is equivalent to Tab~\ref{sps}, but with the ages in Myr}.
\label{pop}
\end{figure*}

The optical properties of NGC\,34 (Mrk\,938, MCG-02-01-032) are
discussed in detail in the recent work of \citet{sch07}. Using
photometry and Lick indices (when spectroscopy was possible) and
comparing them with the models of \citet{bc03} they conclude, that
NGC\,34 has a young stellar disk and supports a rich system of young
star clusters. In fact, as can be observed in Figs.~\ref{popspec} and
\ref{pop} and in Tab.~\ref{sps}, the light at 1.2230\,\mc\ in the 230
central pc of NGC\,34 is also dominated by a stellar population with
solar metallicity and a young-intermediate aged population
($\sim$1\,Gyr). These findings are in full agreement with the fact
that NGC\,34 has in its spectrum a strong 1.1\mc\ CN absorption band
\citep{riffel07} characteristic of stellar populations with that age
(M05).

The colour excess E(B-V) of this source was calculated by
\citet[][hereafter RRP05]{rrp05} using the emission line ratio
Pa$\beta$/Br$\gamma$. They obtain E(B-V)=1.32 mag. As
can be observed, the E(B-V) estimated by means of the
emission lines is  larger than that obtained
from the spectral fitting (Tab.~\ref{sps}). The difference 
can be due to the fact that Pa$\beta$
is strongly affected by absorption lines like Ti\,{\sc i}\,1.2827\mc\
and Ti\,{\sc i}\,1.2851\mc\ as well as Pa$\beta$ itself, as
observed in Fig.~\ref{popspec}. The discrepancy can be also explained
considering that hot ionizing stars can be associated to a dustier region with respect to the cold stellar population
\citep{calzetti94}.

\subsection{NGC\,1614}

This source (Arp\,186, Mrk\,617,II\,Zw\,015) is considered as a
laboratory for studying the evolution of a starburst \citep{alo01}. The
{\it HST}/NIR camera and multiobject spectrometer (NICMOS)
observations reported by \citet{alo01} show deep CO stellar absorption,
tracing a starburst nucleus with a diameter of $\sim$45\,pc,
surrounded by a $\sim$600\,pc diameter ring of extremely luminous
supergiant H{\sc ii} regions.  The stellar population that dominates
the light in the inner $\sim$9\,kpc of this source at 4020\AA\ is
studied by \citet[][hereafter FLL03]{cid03}, using a base of elements
composed by 12 star clusters of different ages and metallicities,
defined by \citet{alex91}. FLL03 find that the dominant stellar
population is 100\,Myr old ($\sim$ 39 \%).  Here we find that the light 
at 1.2230\mc\ in the inner 154\,pc is
dominated by a stellar population of 1\,Gyr age (see Tab.~\ref{sps} and in Fig.~\ref{pop}). 
The dominant 1~Gyr stellar population found in this source as well as in NGC~34 agree with the fact
that the two spectra are very similar.

The difference between our results and those of FLL03 is probably
related to the fact that the aperture used by FLL03
include the light of the younger bulge/arm stars. Interestingly,
we tried a combination of SSPs models 
in the proportions determined by FLL03 in the optical region, but the
resulting synthetic template did not reproduce the observed NIR
spectra of the inner 154\,pc of NGC\,1614.

\citet{pux94} study the dust extinction of this source by means
of hydrogen line ratios. They argue that the NIR line ratios are
inconsistent with a foreground screen of dust obscuring a compact
radiation source, implying a non-uniform distribution of
dust. They also show that if the dust distribution is homogeneous,
then the total extinction is well fitted with an A$_V$=15$\pm$2.5
mag. Using only Pa$\beta$/Br$\gamma$, with a point source model,
\citet{pux94} determine E(B-V)=1.56 (using R$_V$=3.0).  This
E(B-V) value is $\sim$ 2 times larger than that
estimated by RRP05.  This galaxy is the only one of our sample where
the reddening determined by the stellar population fitting
is smaller than that determined by means of emission lines. 
This testifies the complex nature of the dust distribution
in this object. This source should be observed with 2D
spectroscopy in order to map the dust distribution along the galaxy.

\subsection{NGC\,3310}

NGC\,3310 (Arp\,217) is a nearby disk galaxy undergoing intense
starburst activity. It has been the subject of many studies across the
electromagnetic spectrum \citep[see][and references
therein]{elmg02}. This galaxy has a circumnuclear ring of star
formation with angular diameter of 713 to 770 pc \citep[assuming a
diameter between 8$''$ to 12$''$ from][and a distance of 13.24 M\,pc
from NED]{elmg02}.  NGC\,3310 hosts a spectacular starburst, as seen
in numerous H\,{\sc ii} regions. However, it does not show evidence
for even weak O-type stellar populations as seen from the HUT spectrum of
this source \citep{leith02}. The nucleus and six surrounding H\,{\sc
ii} regions, four of which are located at less than 400\,pc from 
the nucleus, were studied in the spectral range between 3600\AA\ and 9600\AA\
by \citet{mgp93}. Here we present the
contributions of the different stellar populations at
1.2230\mc\, in the inner 56\,pc of this object.   As can be observed in
Fig.~\ref{pop} and in Tab.~\ref{sps}, NGC\,3310 displays an intense starburst activity,
with four bursts of star formation. One dominant at 1\,Gyr, which contributes with 30\pp6\% and 
three remarkable bursts at 30\,Myr (13\pp11\%), 50\,Myr (13\pp11\%) and 200\,Myr (12\pp11).

Regarding the reddening, \citet{rrp05} has determined an E(B-V)=0.76
using emission line ratios. As in NGC\,34, the color excess determined
from the stellar population fitting is smaller than that derived from the emission
lines of the gas. This discrepancy is probably due to the fact that the Pa$\beta$ is affected by
absorption lines and/or by the dust distribution, as discussed in the 
case of NGC\,34.

\subsection{NGC\,7714}

NGC\,7714 (Mrk\,538, Arp\,284) has been classified by \citet{weed81}
as a prototypical starburst galaxy.  The optical stellar populations
in the inner $\sim$5\,kpc of this source have also been studied by
FLL03.  They find a 100 Myr old ($\sim$ 33 \%) dominant population.
As for NGC\,1614, we have combined the SSPs in the proportion
calculated by FLL03 but the synthetic template did not
match the observed NIR spectrum in the inner 115\,pc.

Our synthesis results indicates that NGC\,7714
is an active starburst galaxy. As for NGC~3310, three bursts of star
formation were found, a dominant 1\,Gyr (34\pp4\%) old one and two 
minor bursts with ages
30 Myr (10\pp8\%) and 50 Myr (13\pp10\%). The results agree 
with the facts that this galaxy displays a
proeminent 1.1\mc\ CN band \citep{riffel07}  and 
that this source is in merging process due to a
recent off-center collision (100-200\,Myr) with NGC\,7715 \citep{ss03}.

With very little silicate absorption and a temperature of the hottest
dust component of 340~K, NGC~7714 is defined by \citet{brandl04} as
the perfect template for a young, unobscured starburst. However, we
measured an E(B-V)=0.47 from our spectrum, based on hydrogen line
ratios (RRP05). These values agree very well with that determined by
\citet{pux94}. The
E(B-V) determined using the stellar population fitting
is similar to that determined with the emission lines.

In concluding this Section, we call the attention to the fact that our results are not driven by the
CN band. In order to address this point we perform the synthesis for NGC\,7714 excluding the
CN band from our fitting. The age distribution
found in this case is very similar to the one obtained using all absorption lines
\footnote{\begin{small}\renewcommand{\tabcolsep}{0.70mm}\begin{tabular}{lccccccccc}
Age (Gyr)    & 0.001 & 0.01& 0.03  & 0.05& 0.2 & 0.5 & 0.7 & 1   & 13 \\
Cont. (\%) & 7\pp3 &  6\pp4  & 4\pp4 & 14\pp9 & 13\pp10  & 3\pp4 &  6\pp6 & 34\pp5 & 11\pp8\\
\end{tabular}\end{small}}. 

\section{Ionized gas}\label{gas}

After subtracting the stellar population contribution we are left with a
pure emission line spectrum allowing the study of the ionized gas.
We consider that the emission line spectrum of
the galaxies is produced solely by photoionization due to stars. This
assumption is based on X-ray observations reported for the four sources,
which were found to be too weak to be originated by a buried AGN.
In NGC7714, for instance,  the X-ray emission is explained by 
the presence of supernova remnants \citep{weed81}. 
The ROSATASCA observations of NGC\,3310 suggests that the
soft emission is probably a super wind while the nature of the hard 
emission is more uncertain with the likely
origins being X-ray binaries, inverse Compton 
scattering of infrared photons, an AGN or a very hot gas component \citep{zezas98}. However, \citet{liu05} using 
ROSAT High Resolution Imager   
report that the ultraluminous X-ray sources of this galaxy are located 
around knots on spiral arms.  X-ray spectroscopy of NGC\,1614 suggests that
this galaxy may harbor an obscured AGN \citet{risaliti00}, but
VLBI studies with a sensitivity limit of 0.9 mJy do not 
detect a compact radio core in NGC 1614 \citet{hill01}. 
In addition, the three galaxies listed above
are located, in a diagnostic diagram envolving NIR emission line ratios,  in the region correspondig to
SB galaxies \citep[][RRP05]{ara04}. For the last galaxy,  NGC\,34, its nuclear spectrum is 
presently thought to be dominated by a highly obscured starburst, with a likely 
weak AGN contribution \citep{sch07}. 

With the above in mind we compute a grid of models of H{\sc ii} regions in
order to reproduce the observed emission line intensities.  Following
the method of \citet{oli06}, we use the photoionization code
CLOUDY/C07.02 \citep[see][]{cloudy}.
The free input nebular parameters of the models are the abundances
 and ionization parameter $U$.  The nebula was considered as an
expanding sphere, with filling factor $\varepsilon$=0.01 and a
constant electron density of N$_e$\,=\,500\,cm$^{-1}$.  The spectral
energy distribution (SED) used to compute the models are the synthetic
templates derived from the stellar population fitting. We
extrapolate the NIR results from the ultraviolet to the
far-infrared. In this process, we weight the contribution of each
SSPs with the flux contributions listed in
Tab.~\ref{sps}, and sum them. The final SED is very similar
for the four galaxies\footnote{The young stellar populations, 
which ionize the gas, are nearly the same for all galaxies.}. For this reason,
we assume as representative, the resulting template obtained
for NGC\,7714.

\begin{figure}
\centering
\includegraphics[width=8cm]{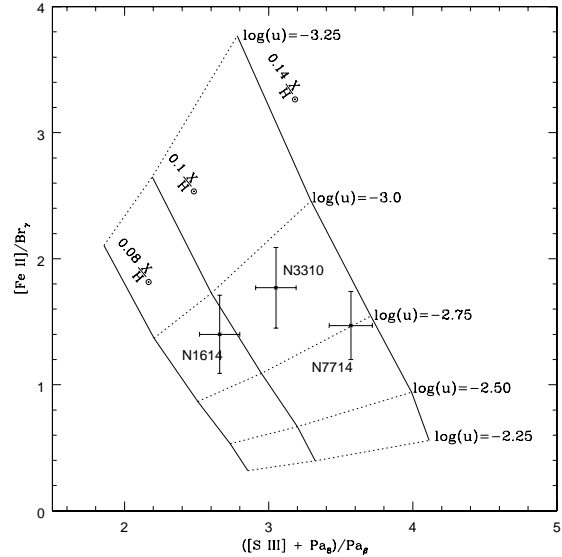}
\caption{Free emission line ratios versus photoionization models.  The grid is computed, while
the data points with error bars represent the observations.  The metallicities and log($U$) are identified.}
\label{sulfe}
\end{figure}

The predicted emission line ratios are in good agreement with 
the observed ones, as can be seen in Tab.~\ref{emint}.  Note that we have
used (\s3\,9531+Pa$\rm _8$)/9069\AA\ instead of the \s3\ line ratio
because the spectral resolution, 20\AA, in this spectral
region does not allow deblending \s3\,9531\AA\ and Pa$\rm _8$
($\lambda$9548\,\AA). In addition, we plot in Fig.~\ref{sulfe}
(\s3\,9531+Pa$\rm _8$)/Pa$\beta$ {\it versus} \fe2/\Br.
The results show that the gas has an ionization parameter in
the range $\rm -3.0 \leq log({\it U})\leq -2.5$, and 
abundances  of S and Fe (X/H)  between
0.08 and 0.14 solar.  The low sulfur abundances  are in
agreement with the findings of \citet{kehrig06}, who
studied the chemical abundances in a sample of 22 H{\sc ii} galaxies
by means of the [S{\sc iii}]~9031,9531\AA\ emission lines. They 
inferred that most
of the objects in their sample have S/H abundances in the range
between 1/20  and  solar. Regarding the
\fe2\, the ratio \fe2\,1.257\,\mc/\Br\ found for our sample
varies from 1.40\pp0.31 to 1.77\pp0.32 (see Tab.~\ref{emint} and Fig.~\ref{sulfe}), similar to
those reported by \citet[][0.65 and 2.70]{simpson96} \footnote{\citet{simpson96} indeed reported 
values of the ratio \fe2\,1.257\,\mc/Pa$\alpha$ in the interval
0.11-0.46 for starburst
galaxies. Assuming case B, Pa$\alpha$/\Br=5.88, we estimated 
that the ratio \fe2\,1.257\,\mc/Pa$\alpha$ translates to \fe2\,1.257\,\mc/\Br\  between 0.65 and 2.70.}.
Models of \citet{colina93} predict ratios between 0.13 and
1.90 for starburst galaxies, well within the range of our observations.

It is interesting to note  that  while the best fitting solutions for the 
stellar populations require SSPs of  solar metallicity, the S and Fe 
abundances of the emission gas are lower than solar. One mechanism that can be
invoked to explain this difference is
interaction among galaxies. Three of the four sources observed  (NGC34, NGC 1614 and NGC 7714)
are strongly interacting objects \citep{thean00,hattori04,dopita02}. 
The gas motions created by the interaction can  significantly alter the 
chemical state of the galaxies \citep{koppen90,dal07},
and modify the usually smooth radial metallicity gradient often found in isolated disk galaxies
\citep{henry99}. Recently, \citet{kewley06} found that the O/H abundance
in the central region of nearby galaxy pairs is systematically lower
than that of isolated objects. These authors suggest that the 
lower metallicity is a  consequence of gas infall caused by the interaction.
In addition, streaming motions of neutral gas (H{\sc i}) towards the 
galaxy center along the spiral arms was  observed in M\,81 \citep{adler96} and 
more recently infall of molecular gas (H$\rm _2$) towards the center of NGC\, 4051 
was reported by \citet{rogemar08}.
Similar motions of molecular gas towards the nucleus along the bar have been  observed in NGC\,4151 
\citep{mundell99}.  In the same way, \citet{mazzuca06} have shown that the star-forming ring
in NGC\,7742 rotates in the opposite direction of the bulge stars, and 
therefore it is made of external material. Therefore, interactions and inflows
 of matter from the outer low metal abundance medium can dilute the nuclear high metallicity interstellar medium.
 
\begin{table}
\begin{scriptsize}
\caption{Properties of the emission line spectrum.  Calculated versus observed values.}
\label{emint}
\begin{tabular}{lccccc}
\noalign{\smallskip}
\hline
\hline
Feature                           & Cal.$^a$& NGC\,34       & NGC\,1614     & NGC\,3310     & NGC\,7714     \\ 
\hline
\noalign{\smallskip}
Pa$\gamma$/Pa$\beta$              &  0.56   &    -          & 0.53$\pm$0.08 & 0.61$\pm$0.05 & 0.52$\pm$0.03   \\
\Br/Pa$\beta$                     &  0.17   &    -          & 0.16$\pm$0.03 & 0.19$\pm$0.02 & 0.18$\pm$0.01   \\
(\s3\,9531+Pa$\rm _8$)/9069\AA    & 2.70    & 2.71$\pm$0.68 & 2.70$\pm$0.13 & 2.67$\pm$0.12 & 2.48$\pm$0.05 \\
(\s3\,9531+Pa$\rm _8$)/Pa$\beta$  &    b    & 1.41$\pm$0.11 & 2.66$\pm$0.14 & 3.05$\pm$0.14 & 3.57$\pm$0.15 \\
\fe2/\Br                          &    b    &	  -	    & 1.40$\pm$0.31 & 1.77$\pm$0.32 & 1.47$\pm$0.27   \\
\noalign{\smallskip}
\hline
\noalign{\smallskip}
\multicolumn{6}{l}{a) Calculated using CLOUDY \citep[see for example][]{cloudy}.} \\
\multicolumn{6}{l}{b) see Fig.~\ref{sulfe}.} \\
\end{tabular}
\end{scriptsize}
\end{table}

\section{Final Remarks}\label{final}

We have analyzed in NIR spectroscopy, from 0.8 to 2.4\,\mc, the stellar
populations in the inner few hundred parsecs of the starburst galaxies
NGC\,34, NGC\,1614, NGC\,3310 and NGC\,7714.  We use a composition of 
simple stellar population models with various ages, from 1 Myr up to 13 Gyr, and 
we also test the effect of metallicity. For the first time, we fit simultaneously 
as much as 15 absorption features in the NIR.
We find that all galaxies host a remarkable fraction of 1\,Gyr stellar populations. 
This result gives support to the fact that the
1.1\mc\,CN band is a powerful tracer of
intermediate age stellar populations \citep{riffel07,maraston05}, though we have 
tested that the ages we derive depend on the multiband fitting rather than on one specific absorption. Our
results are in broad agreement with similar analysis in the optical
region, though we tend to find older ages and a wider age spread. At
least part of this result is due to the fact that the NIR is a region
more suitable to pick up underlying older stellar populations, and to
display the unique absorptions proper to 1\,Gyr-old stellar populations
featured by TP-AGB stars.  Our finding that the central regions of the SB
galaxies contain a substantial fraction of intermediate-age stellar populations
and their prolongued star formation history is very similar to the 
picture drawn by \citet{allard06} and \citet{sarzi07} in the case of central 
star-forming rings, based on optical data. This similarity might support the view
that central star formation often occurs in circumnuclear rings, as was shown 
by the STIS Survey of Nearby Nuclei \citep[SUNNS][]{sarzi05,shields07}.
Analyses based on optical and NIR may be consistent in case of low dust 
obscuration or low contamination from old populations from the bulge region, 
which affect the optical colours. Our NIR based analysis is more robust in 
both respects, because of the whole sampling of the region between $\sim$0.8\,\mc\ and $\sim$2.4\,\mc, 
which includes unique features of intermediate-age such as CN bands.

We found that the metallicity of the stars which dominate the light
at 1.2230\mc, in the inner hundred parsecs of the four galaxies is
most likely solar. The reddening value determined with the whole
spectra is lower than the one determined with the emission lines only.
We found an excellent agreement of the free emission line spectrum with
photoionization models, using as input SED the one derived via the
spectral fitting in the NIR. Due to features
like the 1.1\mc\,CN band, the stellar population analysis in the NIR
seems to allow for a better fine tuning in the age distribution of the galaxy
stellar populations.  The near-IR spectroscopic approach is being
now pursued also in other type of galaxies \citep{silva08}.

\section*{Acknowledgments}

We thank A. L. Chies-Santos, A. C. Krabbe, C. Bonatto, O. L. Dors and
R. S. Nemmen for helpful discussions. R. R. thanks to the Brazilian
funding agency CNPq. MGP thanks to PRONEX/FAPERGS (05/11476).
CM is a Marie Curie Excellence Team Leader and
acknowledges grant MEXT-CT-2006-042754 of the European Community. This
research has been partly supported by the Brazilian 
Brazilian agency CNPq (311476/2006-6) to ARA. This research has made use of
the NASA/IPAC Extragalactic Database (NED) which is operated by the
Jet Propulsion Laboratory, California Institute of Technology, under
contract with the National Aeronautics and Space Administration. The
authors thank the anonymous referee for useful comments about
this manuscript.

\end{document}